%% file: confined_walks.tex
\documentclass[aps,pre,preprint,showpacs,amsmath,amssymb,floatfix,tightenlines,showkeys]{revtex4-2} % PRL -superscriptaddress

\usepackage[pdftex]{graphicx}       % Include figure files
\usepackage{xcolor}
\usepackage{txfonts}        % Times New Roman text & math font
\usepackage{subfig}
\usepackage[normalem]{ulem}

\newcommand{\eref}[1]{Eq.~\eqref{#1}}
\newcommand{\fref}[1]{Fig.~\ref{#1}}
\newcommand{\sref}[1]{Section \ref{#1}}

\usepackage{hyperref} 
\hypersetup{colorlinks,citecolor=blue,filecolor=blue,linkcolor=blue,urlcolor=blue}

\begin{document}

\title{Critical scaling of lattice polymers confined to a box without endpoint restriction}

\author{C. J. Bradly} \email{chris.bradly@unimelb.edu.au}
\author{A. L. Owczarek}\email{owczarek@unimelb.edu.au}
\affiliation{School of Mathematics and Statistics, University of Melbourne, Victoria 3010, Australia}
\date{\today}

% ===========================================================
\begin{abstract}
We study self-avoiding walks on the square lattice restricted to a square box of side $L$ weighted by a length fugacity without restriction of their end points. This is a natural model of a confined polymer in dilute solution such as polymers in mesoscopic pores. The model admits a phase transition between an `empty' phase, where the average length of walks are finite and the density inside large boxes goes to zero, to a `dense' phase, where there is a finite positive density. We prove  various bounds on the free energy and develop a scaling theory for the phase transition based on the standard theory for unconstrained polymers. We compare this model to unrestricted walks and walks that whose endpoints are fixed at the opposite corners of a box, as well as Hamiltonian walks. We use Monte Carlo simulations to verify predicted values for three key exponents: the density exponent $\alpha=1/2$, the finite size crossover exponent $1/\nu=4/3$ and the critical partition function exponent $2-\eta=43/24$. This implies  that the theoretical framework relating them to the unconstrained SAW problem is valid.

\end{abstract}

\maketitle
% ===========================================================
\section{Introduction}

A natural question to ask about lattice polymers is what is the effect of a constraining geometry on the scaling behaviour and physical properties?
Geometric constraints are important for design of polymeric systems in a range of applications \cite{Reisner2005,Luengo1997,Malzahn2016,Torino2016} and have been the subject of numerous molecular simulation studies \cite{Bitsanis1990,Kong1994,Solar2017,Luzhbin2016,Luo2017}.
The mathematical and scaling properties of lattice models of polymers have also had a long history, in particular, the presence and properties of an ordered phase can be significantly altered compared to unconfined polymers.
One class of such studies are those for which the geometric constraint is partially infinite and effectively reduces the dimension of the problem.
The central question in these studies has been to determine the scaling of the number of possible configurations with the size of the restriction.
Early studies for two dimensions looked at walks in a thin strip or slit, effectively a one-dimensional problem where the connective constant is less than unconfined SAWs \cite{Wall1977, Daoud1977, Soteros1988}.
In three dimensions the problem can be restricted to an effectively two-dimensional slab \cite{Whittington1983} or an effectively one-dimensional tube \cite{Benito2018}.
There are also many studies of lattice polymers in a wedge \cite{Guttmann1984,Hammersley1985} including extensions to nonlinear or branched polymers \cite{Soteros1988}.

Another class of models is where the confining geometry is isotropic, rather than partially infinite.
This is physically motivated as a model of a polymer inside a mesoscopic pore which has attracted recent interest \cite{Liu1999,Kipnusu2017}.
The scaling of the number of configurations in such a geometry can still be determined, but for a given system size $L$ there is a maximum length, roughly $n_\text{max} \approx L^d$ for a $d$-dimensional system.
Then the more interesting question is how the system changes from an empty phase where the walks are small to a dense phase where the length of the walks are close to the maximum.
Here we look at polymers confined to a box as an intermediate model between ordinary SAWs on an infinite regular lattice and maximally dense Hamiltonian paths in a box.
The situation where a walk crosses the box -- with endpoints fixed at opposite corners -- has been studied previously \cite{Whittington1990,Madras1995,BousquetMelou2005} with a similar problem introduced earlier by Knuth \cite{Knuth1976}.
Several rigorous results are known about this case, including the scaling dependence on the size of the box and the presence of a critical point between an empty phase and a dense phase.
It is also known that the free energy is strictly bounded by free energies derived from ordinary SAWs and Hamiltonian paths that cross the box \cite{BousquetMelou2005}.
Variations on this model have also been considered, such as fixing the endpoints to opposite sides of the box rather than corners, or allowing more than one mutually-excluding walk within the box \cite{Guttmann2005}.

We wish to completely relax the restriction on the endpoints and look at the weaker condition that the walk is confined to a box of side length $L$, but unrooted with respect to the walls of the box.
Intuitively, our model is intermediate between walks crossing a square and ordinary unconfined SAWs.
All together these results suggest a spectrum of models based on how strict is the geometric confinement.
Given the close relation to these previously studied models, we expect our model to share the same critical phenomena including critical exponents.
Nevertheless, the problem of a single lattice walk confined to a box without tethering, which is in one sense a very natural one to consider in a physical model, has not been laid out in detail, including the scaling behaviour at the critical point between empty and dense phases. So whilst our results may be unsurprising in one sense the natural and fundamental nature of our model motivates the need for its study.

Additionally, our Monte Carlo study reinforces the theoretical framework and exact enumeration analysis in the studies of SAW crossing a square \cite{Whittington1990,Madras1995,BousquetMelou2005} and finite-size scaling in renormalisation group methods \cite{Foster2003a}.
This work is also timely in relation to recent work on the problem of two self-avoiding polygons confined to a box where there is a different phase diagram if the two polygons are separate, or if one polygon is also contained within the other \cite{Rensburg2021}.
Polygon systems where different linking topologies are present are related to polymer knotting \cite{Baiesi2012,DeGennes1984,Delbrueck1962,Koniaris1991}.

In this work we study the model of walks confined to a square box and place this model in the spectrum of models of walks with some confining geometry.
In \sref{sec:Models} we recall known results about ordinary SAWs and SAWs crossing a square and then develop bounds on the free energy of our model.
In \sref{sec:Thermo} we develop the scaling theory of confined SAWs including the finite-size scaling theory near the critical phase transition.
In \sref{sec:MCResults} we describe the Monte Carlo method we use to simulate our model and calculate general thermodynamic properties and critical exponents.
Lastly, in \sref{sec:Trails} we consider the critical behaviour of a similar model using self-avoiding trails confined to a square box.

% ===========================================================
\section{Self-avoiding walk models in confining geometries}
\label{sec:Models}

\subsection{Known results}
\begin{figure}%
	\centering
	\includegraphics[width=\columnwidth]{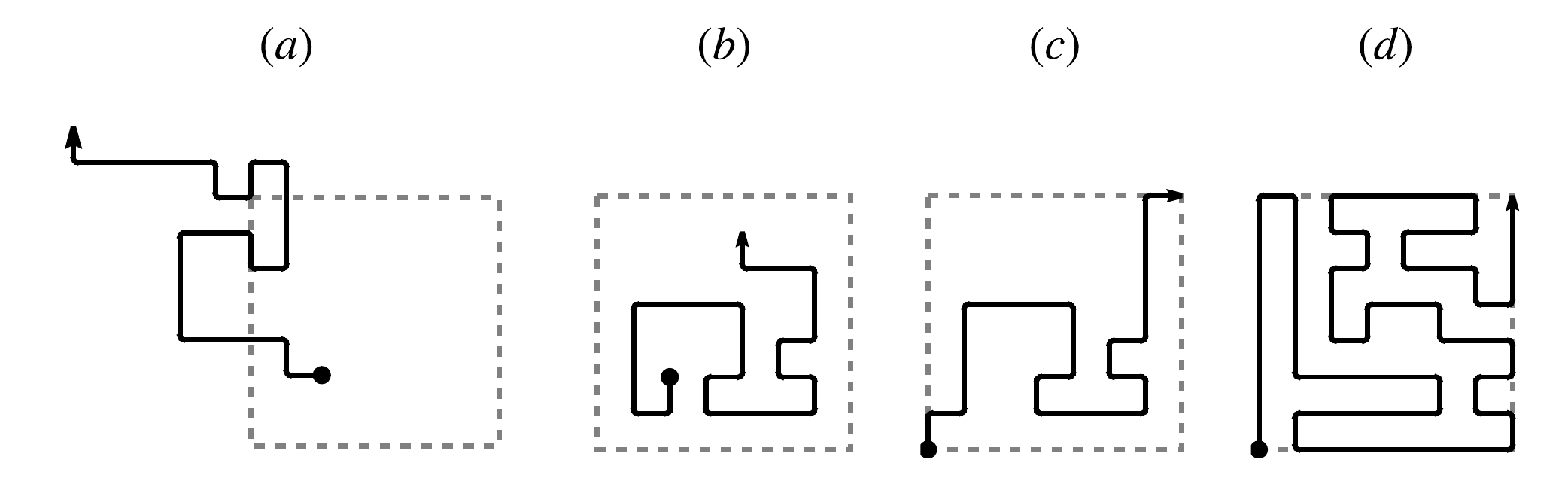}%
	\caption{Example SAWs with increasing degree of confinement to a box of side length $L = 8$. (a) Unconfined, (b) confined to the box (our model), (c) crossing a square and (d) a Hamiltonian path crossing a square.}%
	\label{fig:ExampleSAWsSpectrum}%
\end{figure}

We first recall some results for unconstrained SAWs and SAWs that cross a square.
Example configurations of these models are shown in \fref{fig:ExampleSAWsSpectrum}.

% ===========================================================
% ===========================================================

% ===========================================================
\subsubsection{Unconstrained SAWs}
The canonical model of a polymer in dilute solution is an unconstrained self-avoiding walk on a regular lattice, see \fref{fig:ExampleSAWsSpectrum}(a).
In order to illustrate the comparison to the other models this picture shows an arbitrary box but of course this box has no impact on the properties of unconstrained SAWs.
%In this work we consider two dimensions and the square lattice. 
To avoid ambiguity we consider self-avoiding walks on the square lattice $\mathbb{Z}^2$ with one end at the origin with $n$ steps visiting $n+1$ vertices of the lattice. 
Let the number of such walks be $w(n)$. 
The connective, or growth, constant $\mu$ is defined via the limit \cite{Hammersley1957}
\begin{equation}
	\mu = \lim_{n \to \infty} w(n)^{1/n},
\label{eq:connective_constant}
\end{equation} 
and is specific to each lattice.
For the square lattice $\mu$ has been estimated to high accuracy \cite{Jensen1998,Clisby2013} with the most up-to-date value as $\mu = 2.63815853032790(3)$ \cite{Jacobsen2016}. 
We can define a generating function, or grand canonical partition function, 
\begin{equation}
	G_w(\beta) = \sum_{n=0}^\infty w(n) e^{\beta n},
\end{equation}
which converges for $e^\beta < \frac{1}{\mu} = 0.3790522770(4)$. 
The average length of a walk 
\begin{equation}
	\langle n \rangle  =  \frac{\partial  \log G_w(\beta)}{\partial \beta}
\end{equation}
in the grand canonical ensemble is finite and assuming standard scaling  one can show it diverges as $\beta$ approaches $-\log\mu$ from below.
%as $(-\log \mu - \beta)^{-1}$.

% ===========================================================
\subsubsection{SAWs that cross a square}
We now consider SAWs with end points fixed at two opposing vertices of a square of side $L$ bonds (with $(L+1)$ vertices) and all sites of the walk lie within or on the boundary of the square, see \fref{fig:ExampleSAWsSpectrum}(c). 
This model was first considered in the context of lattice polymers by Whittington and Guttmann \cite{Whittington1990}, and subsequently studied further \cite{Madras1995,BousquetMelou2005}.
Let $s_L(n)$ be the number of walks of length $n$ that are so constrained. 
We note that $ 2L \leq n \leq L^2+2L$ for $L$ even and $ 2L \leq n \leq L^2+2L-1$ for $L$ odd. 
Let us define a partition function by associating a fugacity with the lengths of the SAW via an ``inverse temperature" $ -\infty < \beta < \infty$ as
\begin{equation}
	Z^{(S)}(\beta)_L = \sum_n s_L(n) e^{\beta n}
\end{equation}
and define a finite (reduced) free energy 
\begin{equation}
	f^{(S)}_L (\beta) = \frac{1}{(L+1)^2} \log Z^{(S)}(\beta)_L,
\end{equation}
and the limit 
\begin{equation}
	f^{(S)} (\beta) = \lim_{L \to \infty}   f^{(S)}_L (\beta).
\end{equation}
It has been shown that the limit exists and is finite for all $\beta$ and that 
\begin{equation}
	f^{(S)} (\beta) = 0, \quad\mbox{ for } \beta < -\log \mu\; .
	\label{eq:squarewalks-largenegbeta}
\end{equation}
Moreover, it has been proved that
\begin{equation}
	f^{(S)} (\beta) \leq \log \mu + \beta, \quad\mbox{ for } \beta \geq -\log \mu.
\end{equation}

The number of walks that cross a square $s_L(n) = Z^{(S)}_L(0)$  has been show to grow as
\begin{equation}
	s_L(n) = \lambda_{S}^{L^2 + o(L^2)}
	\label{eq:squarewalks-number}
\end{equation}
where it has been estimated from exact enumeration data that $\lambda_{S} = 1.744550(5)$ \cite{BousquetMelou2005}. 
This gives that 
\begin{equation}
	f^{(S)} (0) = \log \lambda_{S}  
	\label{eq:squarewalks-freeenergyzero}
\end{equation}
and so implies $f^{(S)} (0) > 0$.

% ===========================================================   
\subsubsection{Hamiltonian walks in a square}
An important subset of walks that cross a square are those that are Hamiltonian walks, i.e~they visit every vertex of the square, see \fref{fig:ExampleSAWsSpectrum}(d). 
If the number of such walks is $h_L$ then the limit
\begin{equation}
	\mu_H = \lim_{L\to \infty} h_L^{1/L^2}
\end{equation}
exists and has been estimated as $\mu_H  = 1.472801(1)$ \cite{BousquetMelou2005}. Note that for walks to visit every vertex the side of the square must be even in length. It is also worth noting that while we consider the subset of Hamiltonian walks that cross a square the growth constant seems to be unchanged if one consider all starting and end points in the square. 
Further, it has been shown that \cite{Whittington1990}
\begin{equation}
	f^{(S)} (\beta)\geq \log \mu_H + \beta.
\end{equation}
Thus these Hamiltonian walks are a lower bound to walks that cross a square.
Together with unconstrained walks we have
\begin{equation}
	\log \mu_H + \beta \leq  f^{(S)} (\beta) \leq  \log \mu + \beta.
\end{equation}
Since \eref{eq:squarewalks-largenegbeta} holds this implies that there is a singularity in $f^{(S)} (\beta)$ at some $\beta_\text{c}$ in $ -\log\mu \leq \beta_\text{c} \leq -\log \mu_H$. 
It is predicted that this singularity actually occurs at $\beta_\text{c} = -\log\mu$, see \fref{fig:SchematicFreeEnergy}.

% ===========================================================
\subsection{SAW in a box: our model}
\label{sec:ConfinedSAWBounds}

Now consider walks on a square lattice confined to a box of side length $L$ steps without restriction of their endpoints, see \fref{fig:ExampleSAWsSpectrum}(b). 
The partition function is defined as 
\begin{equation}
	Z^{(B)}_L(\beta) = \sum_n c_L(n) e^{\beta n},
	\label{eq:Partition}
\end{equation}
where $c_L(n)$ is the number of walks of length $n$ that fit in the box and $e^\beta$ is the fugacity of each step.
The maximum length of a walk is the same as for walks that cross a square but without distinction for $L$ odd or even.
However, the minimum length is $n = 0$ so that the number of walks starts with $c_L(0) = (L+1)^2$.
Due to the nature of the Monte Carlo simulations we discuss later it is useful to also consider the number of walks $\hat{c}_L(n)$ that are unique up to translation so that $\hat{c}_L(0) = 1$ with the corresponding partition function $\hat{Z}^{(B)}_L(\beta)$.
We immediately note that $\hat{c}_L(n)\leq c_L(n) \leq (L+1)^2 \hat{c}_L(n)$ for all $n$ and so $\hat{Z}^{(B)}_L(\beta)\leq Z^{(B)}_L(\beta)\leq (L+1)^2 \hat{Z}^{(B)}_L(\beta)$ for all $\beta$. 

We define the finite size free energy
\begin{equation}
  f^{(B)}_L (\beta) = \frac{1}{(L+1)^2} \log Z^{(B)}_L(\beta),
\end{equation}
and respectively $ \hat{f}^{(B)}_L (\beta)$, so that 
\begin{equation}
	\hat{f}^{(B)}_L (\beta)\leq f^{(B)}_L (\beta) \leq  \hat{f}^{(B)}_L (\beta) + \frac{1}{(L+1)^2}\log (L+1)^2.
\end{equation}
We also define the corresponding limits
\begin{equation}
	f^{(B)} (\beta) = \lim_{L \to \infty}   f^{(B)}_L (\beta)  \quad \mbox{and} \quad \hat{f}^{(B)} (\beta) = \lim_{L \to \infty}   \hat{f}^{(B)}_L (\beta).
	\label{eq:free-energy-limits}
\end{equation}
We assume that these limits exist, but note that it has not been proven that they do exist.
It is immediately clear that if the second limit exists then so does the first and they are equal $f^{(B)} (\beta) = \hat{f}^{(B)} (\beta)$.

We can bound our model by comparison to the known results from the other models we have mentioned.
For a lower bound notice that every walk that crosses a square of side $L$ is a configuration of our model so that 
\begin{equation}
	\hat{c}_L(n) \geq s_L(n),
\end{equation} 
and similar for the partition function $\hat{Z}^{(B)}_L(\beta) \geq Z^{(S)}(\beta)_L$ and free energy $ \hat{f}^{(B)}_L(\beta) \geq f^{(S)}(\beta)_L $.
Then assuming the limits in \eref{eq:free-energy-limits} exist we have the bound
\begin{equation}
	f^{(B)} (\beta) = \hat{f}^{(B)}(\beta) \geq f^{(S)}(\beta) \geq  \log \mu_H + \beta.
\end{equation} 

For an upper bound it is also clear that any confined SAW of length $n$ is a valid unconstrained SAW, but not necessarily vice versa, thus
\begin{equation}
	\hat{c}_L(n) \leq w(n),
\end{equation}
which bounds the partition function
\begin{equation}
	\hat{Z}^{(B)}_L(\beta) = \sum_n \hat{c}_L(n) e^{\beta n} \leq \sum_{n\leq (L+1)^2-1} w(n) e^{\beta n} .
\end{equation}
In the region $\beta < - \log \mu $ we have 
\begin{equation}
	\hat{Z}^{(B)}_L(\beta) = \sum_n \hat{c}_L(n) e^{\beta n} \leq \sum_{n\leq (L+1)^2-1} w(n) e^{\beta n}  \leq  \sum_{n\leq \infty} w(n) e^{\beta n} =G_w(\beta) \leq C(\beta),
	\label{eq:finite-regime-bounds}
\end{equation}
for some constant $C(\beta)$ so that 
%\begin{equation}
	$f^{(B)} (\beta) = \hat{f}^{(B)} (\beta) = 0$ %\mbox{  for }  
	for $\beta < - \log \mu$.
%\end{equation}

Now for more general values of $\beta$ we have 
\begin{equation}
	\hat{Z}^{(B)}_L(\beta) = \sum_n \hat{c}_L(n) e^{\beta n} \leq \sum_{n\leq (L+1)^2-1} w(n) e^{\beta n} \leq w\left((L+1)^2-1\right) \sum_{n \leq (L+1)^2-1}  e^{\beta n},
\end{equation}
and so 
\begin{equation}
	\hat{f}^{(B)} (\beta) \leq \frac{1}{(L+1)^2} \log w\left((L+1)^2-1\right) +  \frac{1}{(L+1)^2} \log \sum_{n \leq (L+1)^2-1}  e^{\beta n}.
\end{equation}
Assuming the limits in \eref{eq:free-energy-limits} exist this gives us
\begin{equation}
	f^{(B)} (\beta)=\hat{f}^{(B)} (\beta) \leq  \log \mu + \beta, \quad\mbox{ for } \beta \geq 0,
\end{equation}
and 
\begin{equation}
	f^{(B)} (\beta)=\hat{f}^{(B)} (\beta) \leq  \log \mu,  \quad\mbox{ for } \beta \leq 0.
\end{equation}
More sharply, we know that 
\begin{equation} 
	w(n) = \mu^{n+o(n)},
\end{equation}
which follows from the definition of the connective constant in \eref{eq:connective_constant} \cite{Hammersley1962}.
Hence, for any $\epsilon > 0$ and large enough $L$, we can bound the partition function
\begin{equation}
	\hat{Z}^{(B)}_L(\beta) \leq \sum_{n \leq L^2-1} w(n) e^{\beta n} \leq A  \left( (\mu + \epsilon) e^\beta \right)^{L^2} + C.
\end{equation}
This leads to a bound for the free energy
\begin{equation}
	f^{(B)} (\beta)=\hat{f}^{(B)} (\beta) \leq  \log \mu + \beta, \quad\mbox{ for } \beta \geq -\log \mu.
\end{equation}

So in conclusion $f^{(B)} (\beta) = \hat{f}^{(B)} (\beta) = 0$ for  $ \beta < - \log \mu$ and assuming the limits in \eref{eq:free-energy-limits} exist for $ \beta \geq -\log \mu$
\begin{equation}
	\log \mu + \beta \geq f^{(B)} (\beta) = \hat{f}^{(B)}(\beta) \geq f^{(S)}(\beta) \geq  \log \mu_H + \beta \quad\mbox{ for } \beta \geq -\log \mu.
\end{equation} 

Now it is expected \cite{BousquetMelou2005,Madras1995} that $f^{(S)}(\beta) > 0$ for $\beta > -\log \mu$ so there is a non-analyticity in both $f^{(S)}(\beta)$ and $f^{(B)} (\beta) = \hat{f}^{(B)}(\beta)$ at $\beta = -\log \mu$. 
It may well be the case that $f^{(B)} (\beta)=f^{(S)}(\beta)$ but we are unable to prove that. 

Finally, let us  define
\begin{equation}
	Z^{(B)}_L(0)  = C \lambda_{B}^{L^2 +o(L^2)},
	\label{eq:LambdaConstantBox}
\end{equation}
to consider the total growth of the number of walks inside a box, where $\lambda_{B}$ is the growth constant for walks in a box, in analogy to $\lambda_{S}$ for walks crossing a square, see Eqs.~(\ref{eq:squarewalks-number},\ref{eq:squarewalks-freeenergyzero}).

\begin{figure}%
	\centering
	\includegraphics[width=0.6\columnwidth]{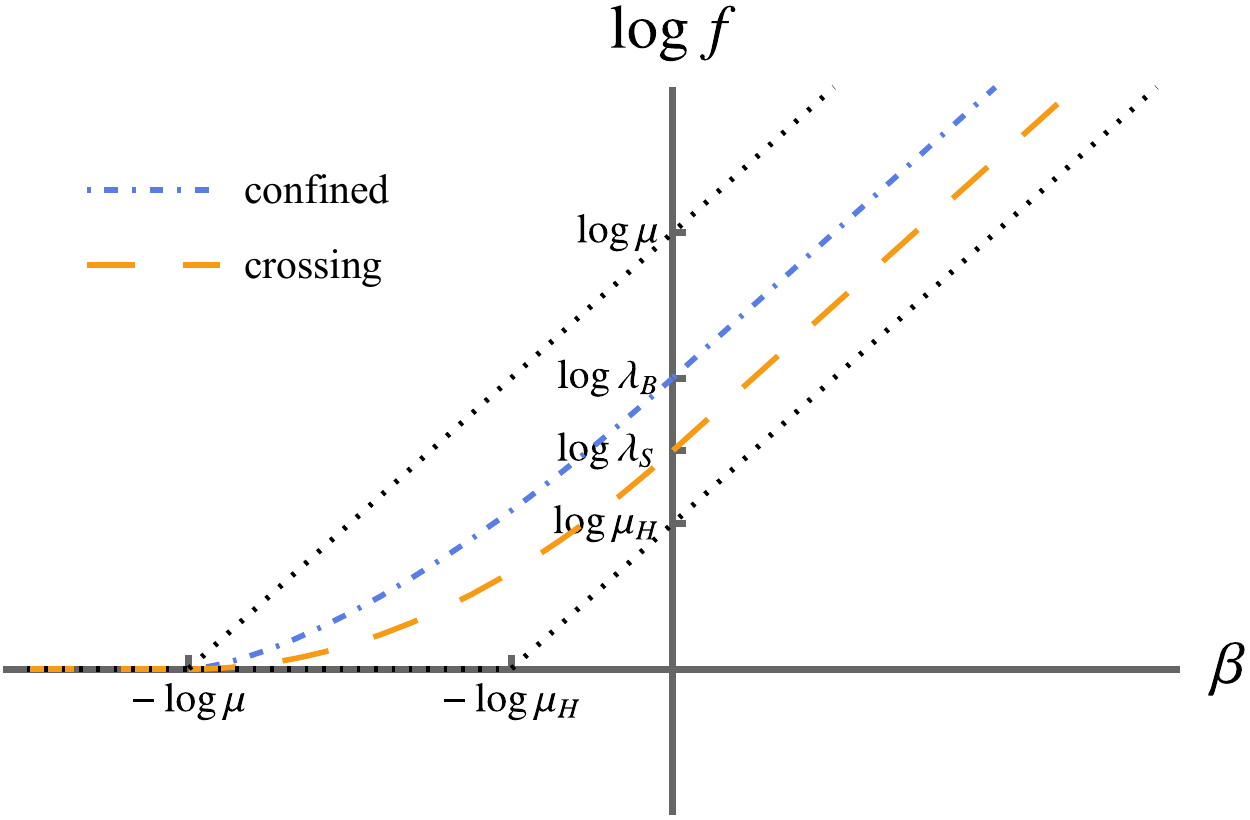}%
	\caption{The free energies of confined SAW models. We do not know that the free energy for our model of confined SAW is strictly greater than that of SAW that cross a square.
	The top and bottom dotted lines mark bounds derived from unconstrained SAWs and Hamiltonian paths, respectively.}%
	\label{fig:SchematicFreeEnergy}%
\end{figure}

Figure \ref{fig:SchematicFreeEnergy} summarises the comparison between the free energies of these models.
The top and bottom bounds are derived from unconstrained SAWs and maximally dense Hamiltonian paths, respectively.

% ===========================================================
% ===========================================================

% ===========================================================
\section{Thermodynamics of walks in a box}
\label{sec:Thermo}

We have seen that the free energy has a singularity which we now infer is a critical point $\beta_\text{c} = -\log\mu$, with $\beta$ serving as the inverse temperature, in line with the model of walks crossing a square \cite{Whittington1990}.
At this critical point there is a phase transition from an empty phase to a dense phase, characterised by the average density changing from zero to a finite value. 
The density is defined as
\begin{equation}
	\rho (\beta) = \frac{\partial  f^{(B)} (\beta)}{\partial \beta},
	\label{eq:Density}
\end{equation}
where we are implicitly assuming that the limits discussed in the previous section can be interchanged with the derivative in \eref{eq:Density}.
Given that the free energy is constant  for  $\beta < -\log\mu$ it is immediately clear that the density is zero in this range and we infer that it is positive and finite for $\beta > -\log\mu$. 
This makes the density a natural order parameter for the transition at $\beta_\text{c} = -\log\mu$.
The system moves from an `empty' phase where the density is zero to a `dense' phase where the density is positive and approaches 1 as $\beta \to \infty$.
The expected length diverges as $\beta$ approaches $\beta_\text{c}$ from below. 
Note for a finite sized system the expected length is still bound by the maximum length possible in the box.

Another indication of the transition occurs by considering the average geometric size of the configurations.
As is standard for studies of models of polymers the relevant quantity is $\langle R^2 \rangle$, where $R$ is the end-to-end distance of the walk, or equivalent metric quantity such as the radius of gyration.
For confined walks we consider the size of the walks relative to the confining box
\begin{equation}
	r_L (\beta)  = \frac{\sqrt{\langle R^2 \rangle}}{L}.
\label{eq:AverageSize}
\end{equation}
with corresponding limit 
$r(\beta) = \lim_{L \to \infty} r_L (\beta)$.
One can see that this is an indicator for the existence of a non-zero density. 
It is zero for  $\beta < -\log \mu$ and is a positive otherwise. We will observe that it is a roughly constant positive value for $\beta > -\log \mu$.

Having described the existence of a critical point we now consider the scaling of thermodynamic quantities near the transition.
Standard critical scaling applies where the free energy is singular with power-law behaviour
\begin{equation}
	f^{(B)}(\beta) \sim |\beta - \beta_\text{c}|^{2-\alpha}, \quad \beta \to \beta_\text{c}^+,
	\label{eq:FreeEnergyTLS}
\end{equation}
and note this exponent is defined only from above $ \beta_\text{c}$ since we have $f^{(B)}(\beta) = 0$ for $\beta < \beta_c$.
This gives us the thermodynamic limit behaviour of the density as
\begin{equation}
	\rho(\beta) \sim |\beta - \beta_\text{c}|^{1-\alpha}, \quad \beta \to \beta_\text{c}^+.
	\label{eq:DensityTLS}
\end{equation}

For a finite system the density of occupied sites in the box is
\begin{equation}
	\rho_L = \frac{\partial f_L}{\partial \beta} = \frac{\langle n \rangle}{(L+1)^2},
\label{eq:FiniteDensity}
\end{equation}
This will be close to zero in the empty phase and non-zero in the dense phase. 
In order to connect to the scaling in the thermodynamic limit we introduce a finite-size scaling ansatz for the (finite-size) density which mirrors the ansatz for walks crossing a square as
\begin{equation}
	\rho_L(\beta) \sim L^q \psi \left( \left(\beta - \beta_\text{c}\right) L^{1/\nu} \right),
	\label{eq:DensityFSSq}
\end{equation}
where $q$ is some leading order exponent specific to the density and $\psi(x)$ is the finite-size scaling function.
The standard finite-size scaling approach is that the scaling variable $x$ is the ratio of the correlation length $\xi$ to the system size $L$.
Near the critical point the correlation length diverges as $\xi \sim |\beta - \beta_\text{c}|^{-\nu}$, with its own characteristic exponent $\nu$ but
for a finite size system, this divergence is also constrained by $\xi \approx L$.
In order for the finite-size scaling ansatz \eref{eq:DensityFSSq} to be consistent with the thermodynamic limit scaling of \eref{eq:DensityTLS} we must choose the relevant scaling variable to be $(\beta - \beta_\text{c}) L^{1/\nu}$.
The exponent $1/\nu$ in this context is the crossover exponent from finite to thermodynamic behaviour.
Then, to produce the free energy singularity in $\beta$, we require $\lim_{L \to \infty} \psi(x) \sim x^{1 - \alpha}$, which will also produce a factor of $L^{(1-\alpha)/\nu}$.
To remove this additional term we therefore need the leading order exponent of $\rho_L$ to be $q = -(1 - \alpha)/\nu$. 
For polymer systems the role of the correlation length is played by the mean-squared end-to-end distance $\langle R^2 \rangle \sim n^{2\nu}$, as described above.
For unconstrained SAW in the canonical ensemble $\nu$ is known exactly to be $3/4$ for non-dense polymers in two dimensions and this has been confirmed numerically to high precision \cite{Clisby2010}. 
In addition to the finite-size scaling ansatz, we assume that near the critical point the typical configurations of confined SAWs are neither dense nor very small and are just long enough to start feeling the walls of the confining box.
We put this as $\sqrt{\langle R^2 \rangle} \approx L$ and thus we have $n \approx L^{1/\nu}$ where the same value of $\nu$ applies as for non-dense unconstrained SAWs.
Comparing this scaling to the direct measure of density $\rho_L = \langle n \rangle/(L+1)^2$ we get the leading order scaling relation $\rho_L \sim L^{-2/3}$.
Matching this to the finite-size scaling ansatz \eref{eq:DensityFSSq} with leading order exponent $q = -(1 - \alpha)/\nu$ we obtain $\alpha = 1/2$.
This value of the critical exponent indicates that the transition between empty and dense phases is a continuous transition. 
So our prediction for two dimensions is that 
\begin{equation}
	\rho_L(\beta) \sim L^{-2/3} \psi\left( \left(\beta - \beta_\text{c}\right) L^{4/3} \right).
	\label{eq:DensityFSSe}
\end{equation}

Given the discussion above one way to characterise this phase transition is to see the scaling of the average length of the walk in the box change as a function of $\beta$:
\begin{equation} 
\langle n \rangle (\beta) \sim
\begin{cases}
	A  &\mbox{ for } \beta < \beta_c\; ,\\
	B \, L^{4/3}  &\mbox{ for } \beta = \beta_c \; ,\\
	C \,  L^2  & \mbox{ for } \beta > \beta_c
\end{cases}
\end{equation}
for constants $A$ and $C$ that are functions of $\beta$ with $A$ diverging like $A(\beta_c -\beta)^{-1}$ as $\beta \rightarrow \beta_c^-$ and $C$ goes to zero like $(\beta - \beta_\text{c})^{1/2} $ as $\beta \rightarrow \beta_c^+$.

The critical behaviour of self-avoiding walks is controlled by two independent exponents (we note above that $\alpha$ and $\nu$ are connected by the hyper-scaling relationship $2-\alpha = d\nu$, where $d$ is the dimension of the system) and are related to two scaling dimensions of the associated conformal field theory \cite{Saleur1987a}. 
For unrestricted walks the second independent scaling dimension appears in the scaling of the canonical partition function through the predicted behaviour $w(n) \sim A \mu^{n} n^{\gamma -1}$ which implies that the grand canonical  generating function diverges as $G_w(\beta) \sim A(\beta_c -\beta)^{-\gamma}$ as $\beta\rightarrow \beta_c^-$. 
Assuming the bound for our partition function $Z_L^{(B)}$ in \eref{eq:finite-regime-bounds} is indicative of the asymptotic behaviour we deduce that
\begin{equation}
Z_L^{(B)}(\beta) \sim \frac{A}{(\beta_c -\beta)^\gamma}.
\end{equation}
In analogy with our scaling ansatz for the density we have
\begin{equation}
Z_L^{(B)}(\beta) \sim L^p \phi \left( \left(\beta - \beta_\text{c}\right) L^{1/\nu} \right),
	\label{eq:PartFnFSSq}
\end{equation}
Similar arguments as above give $p=\gamma/\nu$ and this exponent is  known as $2 - \eta$. 
Hence another scaling relation that we can access is the scaling of the partition function at the critical point in the limit of long walks,
\begin{equation}
	Z_L^{(B)}(\beta_{\text{c}}) \sim B L^{2-\eta},
	\label{eq:PartitionEta}
\end{equation}
where $\eta$ is predicted to have exact value $5/24$ in two dimensions \cite{Nienhuis1982}. 
This scaling form arises from the connection between SAWs and magnetic $O(n)$ vector models where the polymer partition function is derived from the magnetic susceptibility, which scales like $\chi \sim \xi^{2-\eta}$ near the critical point \cite{DeGennes1979}. 
The complete scenario is
\begin{equation}
Z_L^{(B)}(\beta) \sim 
\begin{cases}
	C(\beta)  &\mbox{ for } \beta < \beta_c \; ,\\
	B \, L^{43/24}  &\mbox{ for } \beta = \beta_c \; ,\\	
		\exp\left( f^{(B)}(\beta) \left[L^2 +o\left(L^2\right) \right] \right)  & \mbox{ for } \beta > \beta_c\;.
\end{cases}
\end{equation}
with the constant $C$ being a  non-constant function of $\beta$ and behaves as $ A(\beta_c -\beta)^{-\gamma}$ as $\beta\rightarrow \beta_c^-$.
In this work we test both scaling predictions \eref{eq:DensityFSSe} and \eref{eq:PartitionEta} and verify the three exponents involved.

% ===========================================================
\section{Monte Carlo simulation}
\label{sec:MCResults}

% ===========================================================
\subsection{Simulation method}

\begin{figure}%
	\centering
	\includegraphics[width=0.4\columnwidth]{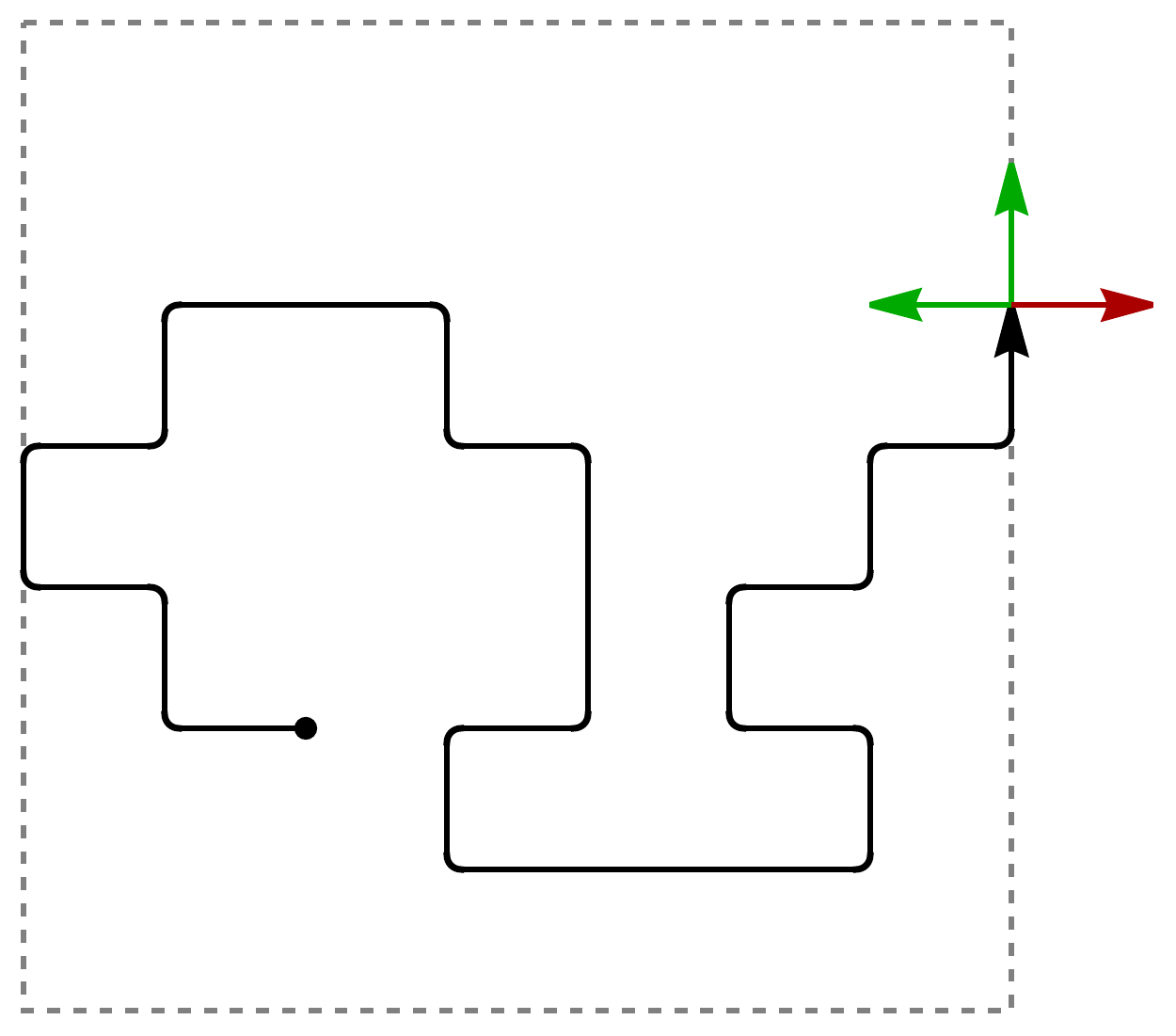}%
	\caption{A sample SAW of length $n = 24$, considered to be confined in a bounding box of side length $L = 7$.
	The possible next steps are shown with arrows; the only restriction is that the step to the right is forbidden if the limit for the simulation was chosen to be $L_\text{max} = 7$.}%
	\label{fig:ExampleConfinedSAW}%
\end{figure}

Simulating confined SAWs is done using the flatPERM algorithm \cite{Prellberg2004} which grows samples step-by-step starting from the origin.
At each step the algorithm uses pruning and enrichment of samples to maintain a flat histogram in the length $n$ and size of the smallest bounding box $L$.
The output of the simulation are estimates of $\hat{c}_L(n)$ from which the partition function and other thermodynamic quantities may be calculated for any temperature $\beta$.
The samples are grown freely up to a predetermined maximum length $n_\text{max}$ with the only restriction being the (predetermined) maximum bounding box size $L_\text{max}$.
Of course the confinement means the predetermined maximum length is restricted by $n_\text{max} \leq  (L_\text{max}+1)^2 - 1$.
Within the simulation we identify the $L$ with the size of the bounding box of a walk so that $L$ is a measured property of each sample.
That is, a walk of length $n$ with bounding box size $L$ whose endpoint lies on the current bounding box can be grown to length $n + 1$ and bounding box size $L + 1$ provided $L < L_\text{max}$.
This is demonstrated in \fref{fig:ExampleConfinedSAW}.
Rather than sampling walks within a predetermined impermeable box, this definition samples walks with arbitrary starting point relative to their bounding box.
In this way it is easier to simulate the number of walks up to translation, i.e.~$\hat{c}_L(n)$.
However, as we showed in \sref{sec:ConfinedSAWBounds} the thermodynamic properties are the same as the model where translations are unique.
The benefit is that the confining box does not need to be specified as a parameter, enabling a more efficient simulation.

% ===========================================================
% ===========================================================

Then the simulation can be configured in one of two ways.
The first is to simply count the number of samples at each $n$ provided they do not exceed a bounding box of size $L_\text{max}$.
In this case $n$ can be sampled up to the largest possible length allowed in a box of size $L_\text{max}$ but no distinction is made for smaller $L$.
This configuration is faithful to the confined SAWs model but only samples a single value of $L = L_\text{max}$ for the purpose of calculating thermodynamic properties.
The second configuration is to record the number of samples at each length $n$ and box size $L$, outputting data for a range of $L$ in a single simulation.
The shortfall of this method is that it does not sample walks with $n < L$, for each $L$; the bounding box of a short straight walk necessarily has the same size as the length of the walk.
However, as we shall see, these missed samples are only significant for the empty phase which we know to be trivial.
In return we can simulate many more samples at larger $L$ by restricting $n_\text{max}$ to less than that allowed by $L_\text{max}$.
This is useful for determining the scaling near the critical point.

% ===========================================================
\begin{figure}%
	\centering
	\includegraphics[width=0.7\columnwidth]{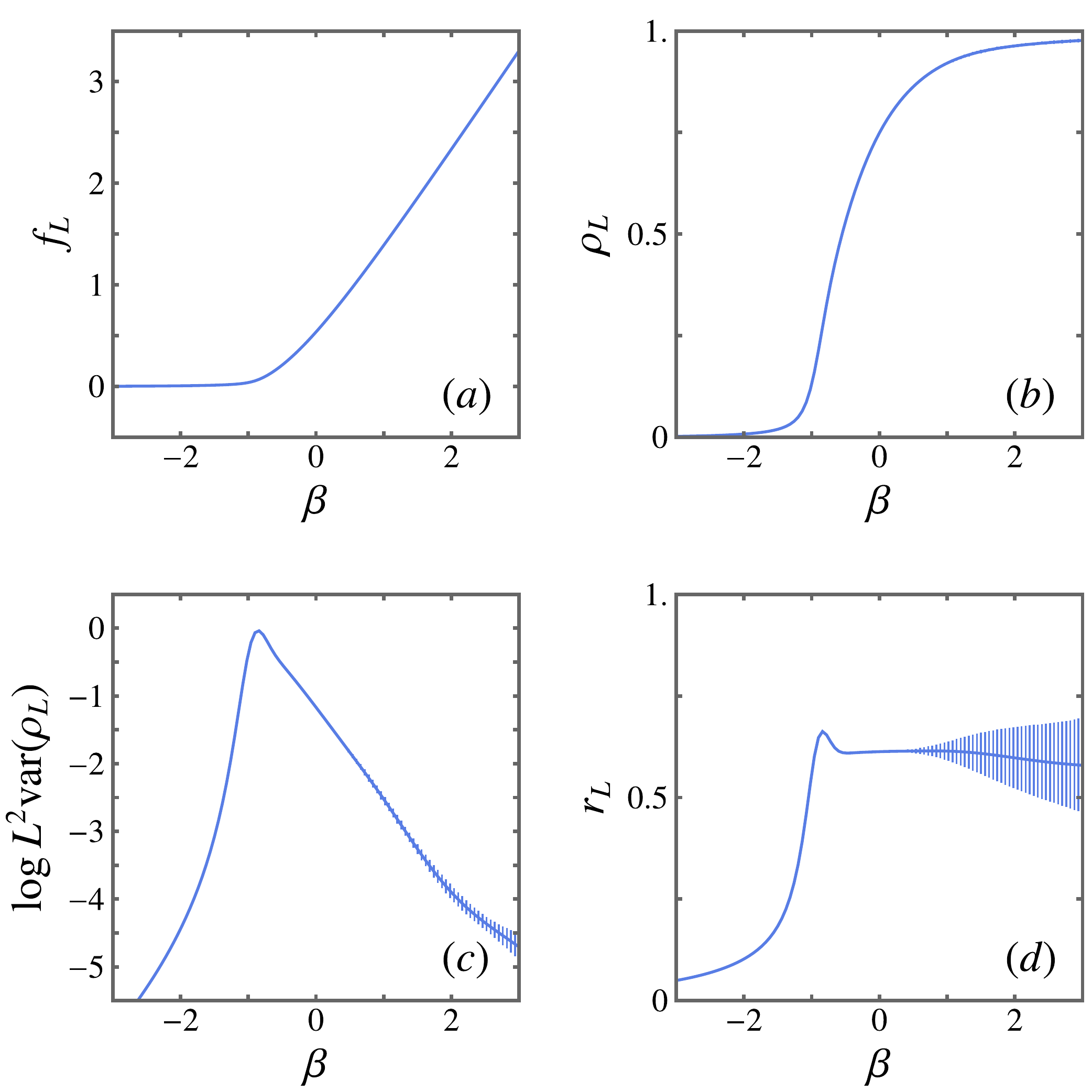}%
	\caption{Thermodynamic quantities for SAWs confined to a box of size $L = 9$. 
	Plots show (a) the free energy $f_L$, (b) the density $\rho_L$, (c) the logarithm of the variance $L^2 \text{var}(\rho_L)$ , and (d) the average size $r$.}%
	\label{fig:GeneralThermoWalks}%
\end{figure}
% ===========================================================

% ===========================================================
\subsection{General properties of confined SAWs}
Before considering the critical behaviour more closely, we verify the general thermodynamic properties of confined SAWs with the simulation of a small system with $L_\text{max} = 9$ and $n_\text{max} = 99$.
This simulation used the first configuration to sample the entire range of lengths for a box of size $L_\text{max} = 9$.
In \fref{fig:GeneralThermoWalks} we plot several thermodynamic quantities calculated as a function of the parameter $\beta$, namely (a) the free energy $f_L$, (b) the density $\rho_L$, (c) the variance of the density $L^2 \text{var}(\rho_L)$, and (d) the average size of the walks $r_L$.
For results from a finite-size simulation we calculate the density directly as $\rho_L = \langle n \rangle/L^2$ and the variance as $L^2\text{var}(\rho_L) = \langle n^2 \rangle - \langle n \rangle^2 $, where the average $\langle \cdot \rangle$ is calculated from the estimates of $c_L(n)$.
For the average size $r_L$, the metric quantity $\langle R^2 \rangle$ is calculated within the simulation and is a separate output to the estimates of $c_L(n)$.
The empty phase, $\beta < \beta_\text{c}$, has $f_L \approx 0$, $\rho_L \approx 0$ and $r_L \approx 0$.
The dense phase, $\beta > \beta_\text{c}$, has $f_L \sim \beta$, $0< \rho_L \leq 1$ and $r_L \approx \text{constant}$.
The critical point $\beta_\text{c} \approx -0.97$ separates the two phases and is most clearly marked by a peak in the variance of $\rho_L$.
The small peak in $r_L$ near the critical point is a finite-size effect that diminishes as $L$ is increased.
The increase in uncertainty in $r_L$ in the dense phase as $\beta$ increases
is due to the difficulty in sampling very dense configurations, even for a relatively small box.

% ===========================================================
\subsection{Critical scaling of confined SAWs}

To verify the scaling relations for the critical point we need data for a larger range of $L$.
FlatPERM has the advantage that a single simulation outputs data for all $n$ and $L$ up to some predetermined maximum, but since the longest possible samples have lengths $L^2$ this can quickly become unmanageable.
However, as we have already pointed out, near the critical point the system is dominated by configurations with length $n \sim L^{4/3}$.
Thus we can simulate larger $L$ by restricting $n$ in the second configuration, described above.
In practice, we simulate up to box size $L_\text{max} = 128$ with maximum walk length $n_\text{max} = 1280$ to cover the samples that are most important to critical properties of this system.

\begin{figure}%
	\centering
	\subfloat{\includegraphics[width=0.35\columnwidth]{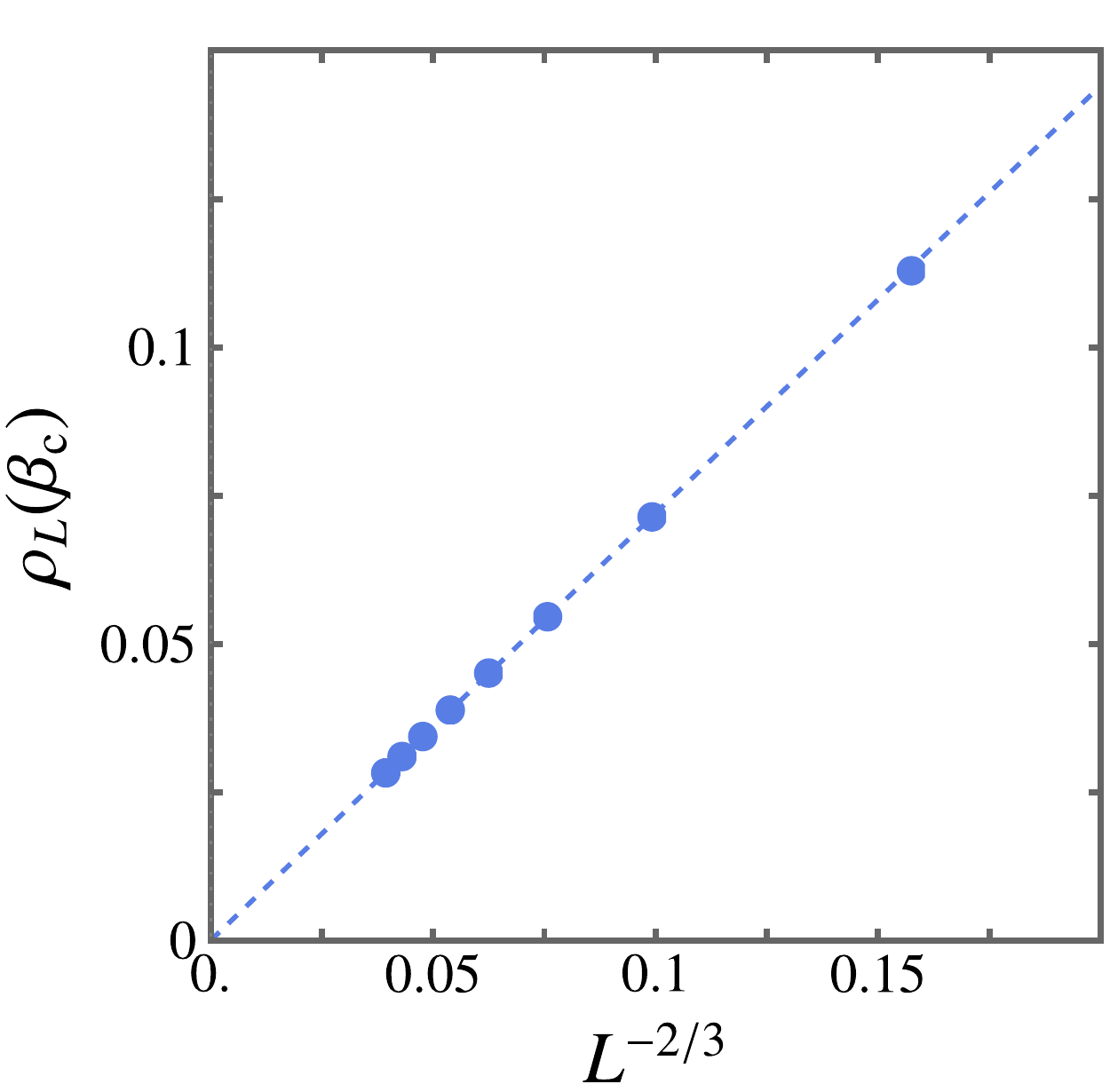}} \quad
	\subfloat{\includegraphics[width=0.35\columnwidth]{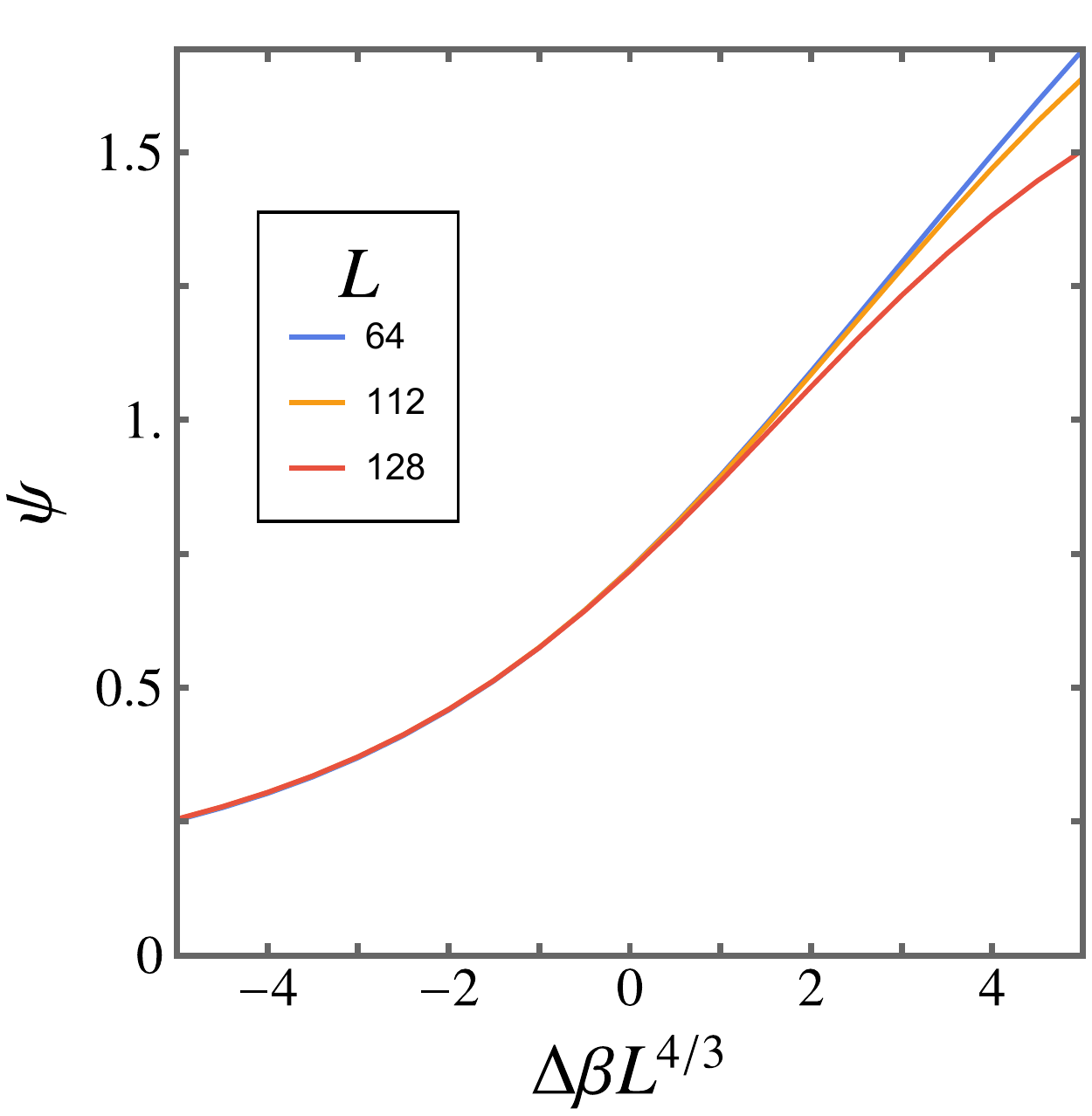}}
	\caption{(a) The critical density $\rho_L(\beta_\text{c})$ plotted against the expected scaling $L^{-2/3}$ and (b) the scaling function $\psi(x)$ for confined SAWs.}%
	\label{fig:DensityAndScalingWalks}%
\end{figure}

Firstly, we show in \fref{fig:DensityAndScalingWalks}(a) the density calculated at the critical point $\beta_\text{c} = -0.970$ over a range in $L$.
The leading order behaviour is clearly consistent with $L^{-2/3}$.
More precisely, we fit the data to \eref{eq:DensityFSSq} assuming the scaling function is a constant at $\beta = \beta_\text{c}$ and $\nu = 3/4$.
This is shown by the dashed line and yields the critical exponent $\alpha = 0.4996(8)$ matching the expected value $1/2$ accurately.
Note that although we plot the density against $L^{-2/3}$ this power law is not assumed for the estimate of $\alpha$.

Next we consider the crossover exponent in the scaling variable $x = \Delta\beta L^{1/\nu}$.
From \eref{eq:DensityFSSq} the derivative of the density with respect to $\beta$ will extract a factor $L^{1/\nu}$ which can be isolated by dividing by $\rho_L$.
But the derivative of the density is the second derivative of the free energy, or simply the variance of the density.
Thus the ratio $L^2 \text{var}(n)/\rho_L$, evaluated at the critical point, where the scaling function and its derivative are constant, will have leading order $L^{1/\nu}$.
A simple fit of this ratio to a power law yields $\nu = 0.756(4)$, consistent with the known value $\nu = 3/4$.

% ===========================================================
% ===========================================================

Having verified the exponents $\alpha$ and $1/\nu$ we can now calculate the scaling function as $\psi(x) = L^{2/3}\rho_L \left( L^{-4/3}x + \beta_\text{c} \right)$.
In \fref{fig:DensityAndScalingWalks}(b) we plot this quantity for several values of $L$.
The match is excellent for $x \lesssim 1$ and indicates that the scaling is correct.
The curves only diverge for larger $x$ and large $L$ where the density calculated from the simulation data is less accurate, mainly due to the cutoff of lengths below $L^2$.
The scaling function calculated for smaller $L$ all lie on top of the $L = 64$ line.

\begin{figure}%
	\centering
	\includegraphics[width=0.35\columnwidth]{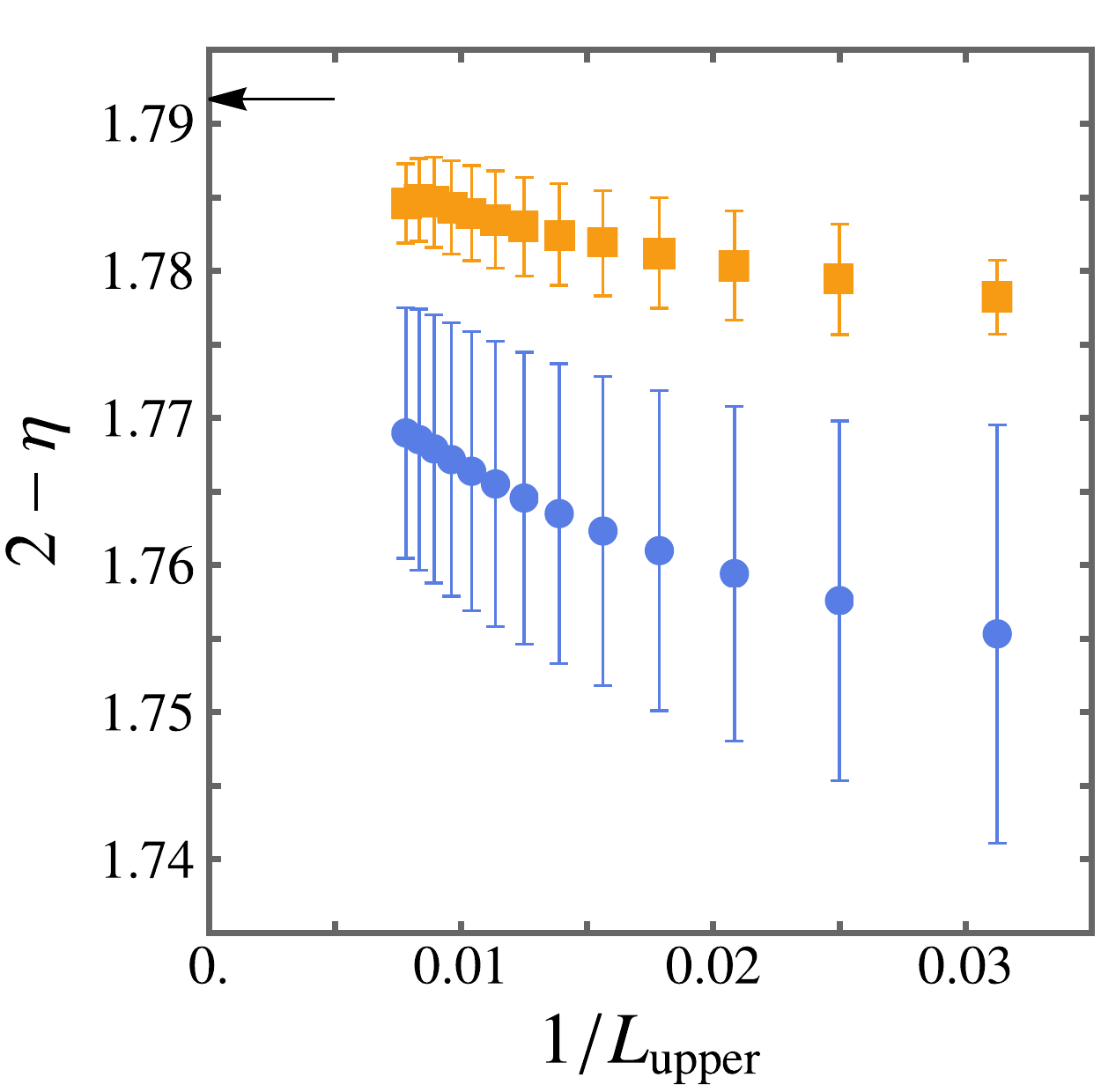}%
	\caption{The critical exponent of the confined SAW partition function $Z_L(\beta_\text{c})$ versus the upper bound of the range of $L$ values used to fit the data, with (top) and without (bottom) a correction-to-scaling term.}%
	\label{fig:PartitionEtaWalks}%
\end{figure}

Finally, we can also estimate the critical exponent $\eta$ by fitting our data to \eref{eq:PartitionEta}. 
However, simply fitting to the power law $L^{2-\eta}$ yields a poor estimate for $\eta$.
Series analysis of the generating function of a lattice walk model which predicts large $n$ asymptotic behaviour of the coefficients near the critical point to have multiple additional terms, both analytic and terms with correction-to-scaling exponents \cite{Conway1996}.
For the range of $L$ values we consider here the additional terms can be significant but we cannot distinguish which correction-to-scaling terms are most important and so we include an additional $1/L$ term to broadly capture this effect.
Additionally, to demonstrate the scaling as $L$ increases we repeat the fit over different ranges of $L$, capped by the upper bound $L_\text{upper}$.
The results are shown in \fref{fig:PartitionEtaWalks} where the top set was calculated with the additional $1/L$ term and the bottom set without.
The known exact value using $2-\eta = 43/24=1.791\dot{6}$ marked with an arrow; our best estimate is $2-\eta = 1.785(3)$.
The addition of a $1/L$ term has a marked effect, both producing estimates closer to the known value and reducing the error bars, but still falls short of the exact value.
This is to be expected given that the range of $L$ is relatively small for estimating the behaviour of very long polymers.
However, our estimates appear to trend towards the known value of $\eta$ and suggest that data for larger box sizes could be useful for extrapolating to the large $L$ limit.

% ===========================================================
% ===========================================================

% ===========================================================
\subsection{Confined self-avoiding trails}
\label{sec:Trails}

While self-avoiding walks are the canonical model for dilute lattice polymers, we can also consider a similar model called the self-avoiding trail (SAT).
In this model the walk is allowed to visit the same site multiple times, but cannot revisit a previous bond or step.
This restriction still represents the excluded-volume effect needed to model polymers but can have different mathematical properties.
SATs are often used as an alternative to SAWs for modelling polymer collapse induced by a monomer-monomer interaction representing solvent quality \cite{Owczarek1995,Owczarek2007,Foster2009}.
Although collapse is a different phenomenon it is also a transition between two phases typified by a difference in metric size.
Thus it is of interest to consider SATs in the context of confined polymers.

Much of the background and scaling analysis we outlined in \sref{sec:Models} and \sref{sec:Thermo} applies to trails as well.
The connective constant for SATs on the square lattice is slightly different than for SAWs but has also been estimated to high accuracy as $\mu_\text{SAT} = 2.72062(6)$ \cite{Conway1993, Guim1997}, thus we expect $\beta_\text{c} = -1.0008605(27)$ for confined SATs.
Using this as the critical point we estimate the critical scaling exponents for density in the same ways as for confined SAWs.
We simulated SATs confined to a box with $L_\text{max} = 128$ and $n_\text{max} = 1280$.
This is the same system size as the SAW simulations, but we note that for trails of the same length $n$ are less dense in the box since a smaller fraction of sites can be occupied due to double occupation; the maximum density is $\rho_L = 2$ using the same definition as for SAWs.
However, these simulation parameters are still large enough to cover the critical behaviour of confined SATs.

\begin{figure}%
	\centering
	\subfloat{\includegraphics[width=0.35\columnwidth]{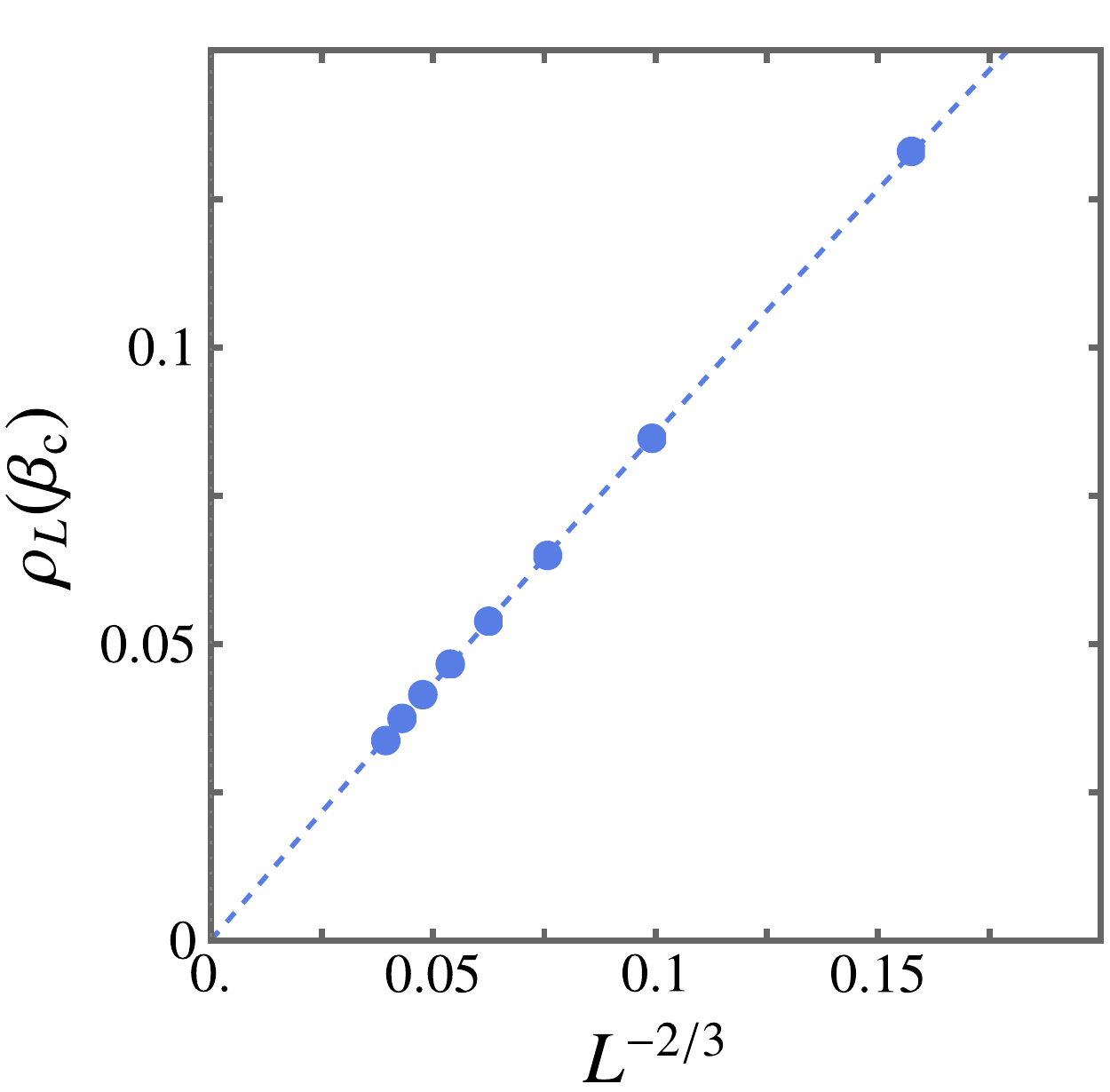}}
	\qquad
	\subfloat{\includegraphics[width=0.35\columnwidth]{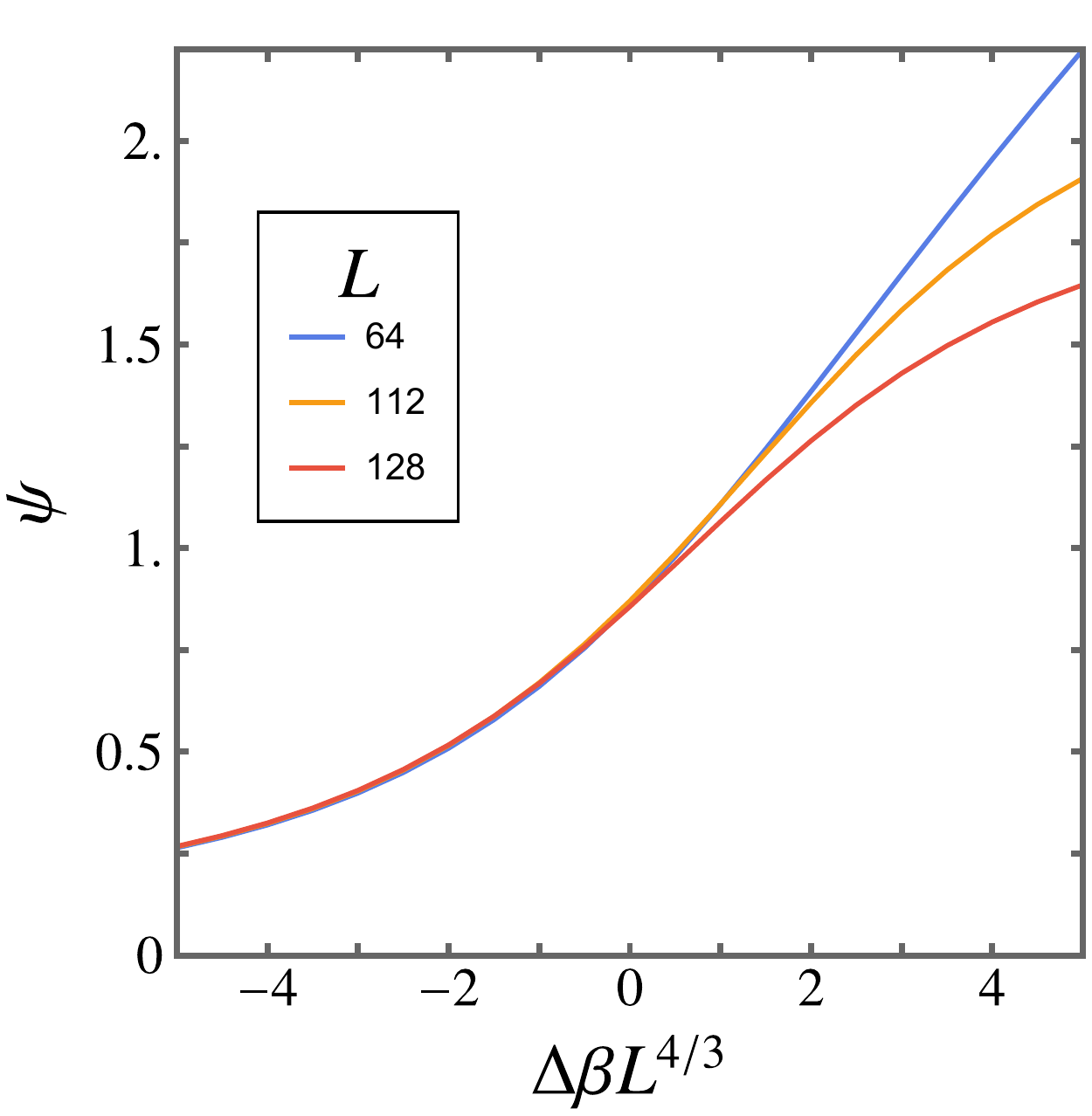}}
	\caption{(a) The critical density $\rho_L(\beta_\text{c})$ plotted against the expected scaling $L^{-2/3}$ and (b) the scaling function $\psi(x)$ for confined SATs.}%
	\label{fig:DensityAndScalingTrails}%
\end{figure}

In \fref{fig:DensityAndScalingTrails}(a) we plot the density of confined SATs at the critical point.
We also show the fits to the density using the ansatz of \eref{eq:DensityFSSe} from which we estimate $\alpha = 0.498(4)$.
This is also consistent with the expected value of $\alpha = 1/2$ although the error is slightly larger than for SAWs.
Finally, we also calculate the scaling function for confined SATs, shown in \fref{fig:DensityAndScalingTrails}(b).
Comparison with \fref{fig:DensityAndScalingWalks}(b) indicates that the scaling function for SATs is similar to that for SAWs, but is not valid over the same range of $x > 0$; the divergence of curves for larger $L$ happens sooner for SATs compared to SAWs.

% ===========================================================
% ===========================================================

% ===========================================================
\section{Conclusion}
\label{sec:Conclusion}

We have studied a model of SAWs confined to a box on the square lattice without restriction on the location of its endpoints.
We were able to bound the free energy of our model in relation to other models of SAWs with similar confining geometry.
We find that our model is intermediate to the canonical model of unconstrained SAWs and the model of SAWs that cross the box with endpoints fixed to opposing corners.

SAWs confined to a box have two phases, empty and dense, and a critical phase transition between them.
The critical point is related to the value of the connective constant for unconstrained SAWs.
We have developed the scaling theory for the thermodynamic properties of our model, including finite-size scaling theory near the critical point and expected critical exponents.
Monte Carlo simulation was used to verify both the general thermodynamic properties of our model, and estimate the values of these critical points.
Our numerical results confirm that the density scales with leading order $L^{-2/3}$ indicating that $\alpha = 1/2$ and thus the transition is continuous.
We predict that other similar systems, such as confined linked polygons \cite{Rensburg2021}, should have this same scaling in the regime where one polymer dominates. In particular, the density exponent should be $\alpha=1/2$. 
We also verified that the crossover exponent is $1/\nu$ where $\nu = 3/4$ is the standard Flory exponent for polymers in two dimensions.
In the process we were also able to isolate the finite-size scaling function in the vicinity of the critical point. 
Lastly, our data is consistent with the leading scaling exponent of the partition function at the critical point $2-\eta=43/24$, although corrections to scaling are significant.
These critical exponents match the exponents for the unconfined SAW model, which can be considered the limiting large-$L$ case.

We applied the same analysis to the similar model of self-avoiding trails confined to a box, finding that there are additional terms to account for and larger numerical errors.
Nonetheless, we expect that the SAT model likely has the same critical exponents which would put the empty-dense transition for confined SATs in the same universality class as confined SAWs. 

Finally, we remark on the scaling of the our confined SAWs model.
Analogous to the model of walks crossing a square, one defines the growth constant for the total number of SAWs in a box $\lambda_{B} \equiv e^{f^{(B)}(0)}$, see \eref{eq:LambdaConstantBox}.
A numerical estimate of $\lambda_{B}$ compared to $\lambda_{S} = 1.744550(5)$ could indicate if the model of walks crossing a square is a strict lower bound to general confined walks or that the two models share a similar free energy.
Unfortunately we were unable to get good enough data using our methods to estimate $\log\lambda_{B} = f^{(B)}(0)$.
At $\beta = 0$ dense configurations are more important than near the critical point and these are hard to sample for the larger values of $L$ that would be required to provide a good estimate.
Other methods such as exact enumeration would be helpful here and allow a better comparison to the model of SAWs crossing a square.

%%%%%%%%%%%%%%%%%%%%%%%%%%%%%%%%%%%%%%%%%%%%%%%%%%%%%%%%%%%%%
%%%%%%%%%%%%%%%%%%%%%%%%%%%%%%%%%%%%%%%%%%%%%%%%%%%%%%%%%%%%%
\begin{acknowledgments}
Financial support from the Australian Research Council via its Discovery Projects scheme (DP160103562) is gratefully acknowledged by the authors.
\end{acknowledgments}

% ===========================================================
% ===========================================================
%\bibliography{../../Manuscripts/polymers_master}{}
%\bibliography{../../polymers_master}{}
%\bibliography{../polymers_master}{}
\input{confined_walks.bbl}
\end{document}